\documentclass[10pt]{iopart}

\usepackage{iopams}
\usepackage{setstack}
\usepackage{graphicx}
\usepackage[caption = false]{subfig}
\usepackage{color}
\graphicspath{{Figures/}}
\usepackage{citesort}
\usepackage[numbers,sort&compress]{natbib}
\usepackage{hyperref}
\usepackage{ulem}

\begin{document}

\title[]{Entropy production at criticality in a nonequilibrium Potts model}

\author{Thomas Martynec$^1$, Sabine H.L. Klapp$^1$, Sarah A.M. Loos$^1$}

\address{$^1$Institute for Theoretical Physics, Technische Universit\"at Berlin,
Hardenbergstr. 36, D-10623, Berlin, Germany}
\ead{martynec@tu-berlin.de}

\begin{abstract}
Understanding nonequilibrium systems and the consequences of irreversibility for the system's behavior as compared to the equilibrium case, is a fundamental question in statistical physics. Here, we investigate two types of nonequilbrium phase transitions, a second-order and an infinite-order phase transition, in a prototypical $q$-state vector Potts model which is driven out of equilibrium by coupling the spins to heat baths at two different temperatures. We discuss the behavior of the quantities that are typically considered in the vicinity of (equilibrium) phase transitions, like the specific heat, and moreover investigate the behavior of the entropy production (EP), which directly quantifies the irreversibility of the process. For the second-order phase transition, we show that the universality class remains the same as in equilibrium. Further, the derivative of the EP rate with respect to the temperature diverges with a power-law at the critical point, but displays a non-universal critical exponent, which depends on the temperature difference, i.e., the strength of the driving. For the infinite-order transition, the derivative of the EP exhibits a maximum in the disordered phase, similar to the specific heat. However, in contrast to the specific heat, whose maximum is independent of the strength of the driving, the maximum of the derivative of the EP grows with increasing temperature difference. We also consider entropy fluctuations and find that their skewness increases with the driving strength, in both cases, in the vicinity of the second-order transition, as well as around the infinite-order transition.
\end{abstract}

\maketitle

\section{Introduction}

Phase transitions are ubiquitous in nature and generally occur in equilibrium as well as nonequilibrium systems. In either case, the transition is due to internal interactions and often goes along with the breaking of spatial symmetries as a reaction to the variation of a control parameter below its critical value, detectable by the emergence of an appropriate order parameter. Phenomena like the occurrence of phase transitions in one spatial dimension are solely observed in nonequilibrium systems~\cite{Marro_1999,Hinrichsen_Advances,Dickman_1991,Schmittmann_1995,Menyhard_1994, Menyhard_1995,Menyhard_1996,Hinrichsen_1997,Castellano_2000,Hinrichsen_2000,Lipowski_2000,Luebeck_2004,Holme_2006}. In contrast, other properties related to phase transitions are identical and thus do not allow perceiving whether a system is in a state of thermal equilibrium or not. In equilibrium, it is well established that continuous (second-order) phase transitions are accompanied by power-law divergences of multiple measurable quantities, such as the magnetic susceptibility or the spin--spin correlation length (for Ising-like models) and there are already numerous examples for nonequilibrium systems that can be characterized in this manner as well~\cite{Hagen_Book,Oliveira_1993,Blote_1990,Hasenbusch_1999,Deng_2003,Hasenbusch_2010,Hasenbusch_2001}. However, a general theory for nonequilibrium phase transitions is still missing, and it is not \textit{per se} clear whether the critical exponents of a system stay the same (i.e., the system remains in the same universality class) when driven away from equilibrium. 

To analyze nonequilibrium systems there exists another, yet almost completely unconnected, tool, that is, the entropy production (EP). This quantity is strictly positive for nonequilibrium and exactly zero for equilibrium systems. The total EP is a fundamental quantity of great importance in statistical physics that already plays a central role in (stochastic) thermodynamics and information theory, and it is known to fulfill various laws, including the famous fluctuation theorems \cite{Kurchan_1998,Crooks_1999,Seifert_2005,Seifert_2012,Zhang_2012,Vandenbroeck_2015,Herpich_2020}. Moreover, the EP can be defined in a very general manner (not system-specific) and is, in principle, a meaningful quantity for any complex system. This is because it solely depends on (state and transition) probabilities, and does not rely on concepts like energy or temperature. Thus, it can be defined and calculated also, e.g., in non--physical systems, like social dynamics, or opinion formation, which are at the same time known to undergo phase transitions~\cite{Hinrichsen_2001,Dasilva_2020}. For these reasons it is tempting to investigate this quantity with regard to critical behavior in nonequilibrium systems.

A few recent studies have already started to investigate the behavior of total EP around criticality in different lattice-based models by means of Monte-Carlo simulations and mean field theory \cite{Tome_2012,Noa_2019}. For example, for a one-dimensional KPZ interface growth model~\cite{Barato_2010,Barato_2012} the rate of EP around the critical point of a first-order phase transition was calculated. Furthermore, there are several studies considering spin systems with up-down ($Z_2$) symmetry. These include an interacting lattice gas model in contact with two heat and particle reservoirs \cite{Tome_2012}, the majority vote model \cite{Crochik_2005,Silva_2020}, and an Ising model externally driven by an deterministically oscillating magnetic field \cite{Zhang_2016}. In all cases, the EP rate was either found to jump or display an inflection point at criticality, such that its first derivative with respect to an appropriate control parameter exhibits marked behavior around the critical point. In particular, the derivative shows a discontinuity or a power-law divergence, bearing a resemblance with the susceptibility, the spin--spin correlation length and the specific heat. Of special interest for the present work is a study considering a variant of the square lattice Ising model with nearest-neighbor interactions, whose spins are in contact with two heat baths at temperatures $T_{1}$ and $T_{2}$ in a checkerboard arrangement~\cite{Tome_1991,Barbosa_2018}. In this study, the derivative of the EP was found to diverge around the second-order phase transition with the \textit{same} critical exponent as the specific heat~\cite{Barbosa_2018}. This raises the question whether these diverging quantities generally behave alike at criticality. 

In the present paper, we aim to generalize the previous findings to spin symmetries different from $Z_2$, and to other types of phase transitions. To this end, we consider a nonequilibrium $q$-state vector Potts model around criticality. Depending on the value of $q$, the model displays either a second-order phase transition from a paramagnetic (PM) to a ferromagnetic (FM) phase or an infinite-order phase transition \cite{Surungan_2019,Sun_2019,Ueda_2020} from a PM to a quasi long-range ordered Berezinskii–Kosterlitz–Thouless (BKT) phase. We investigate the model in the vicinity of both types of phase transitions under nonequilibrium conditions. In particular, we couple the spins to heat baths at two different temperatures $T_1$ and $T_2$ in such a way, that all nearest-neighbors of each spin are in contact only with spins coupled to the respective other heat bath. Hence, two sublattices are formed in a checkerboard arrangement similar to~\cite{Tome_1991,Barbosa_2018}. As a consequence of the two involved temperatures, a net heat flow from the hotter to the colder heat reservoir is induced. This setup drives the system in a nonequilibrium steady state where it constantly produces entropy, which is exported to the environment. We investigate the system numerically using Monte Carlo simulations with Glauber dynamics. To deepen our understanding of nonequilibrium phase transtions, we study the EP and its fluctuations, as well as standard quantities such as the magnetization. Using the finite-size scaling technique \cite{Binder_1981,Brezin_1982,Cardy_1984,Chayes_1986} to carefully analyze the critical behavior based on numerical data, we dedicate a detailed analysis to the question whether the specific heat and the derivative of the EP behave alike at criticality for both types of phase transitions. 

Our analysis reveals that the derivative of the EP rate with respect to temperature in the PM disordered phase of the $4$-state vector Potts model shares indeed similarities with the specific heat. Specifically, it shows power-law behavior as function of the distance from criticality, however, in contrast to the specific heat, with non-universal scaling exponent. Additionally, its maximum value diverges at criticality and also shows power-law behavior as function of system size with a non-universal scaling exponent. In contrast, for the XY model with $q \rightarrow \infty$, the EP rate resembles the behavior of the specific heat. Both quantities do not diverge in the vicinity of the transition from the PM to the BKT phase. 

\section{Modeling and simulation details}
\label{sec: model}
\subsection{The $q$-state vector Potts model}

The Hamiltonian of the $q$-state vector Potts model (or the $q$-state clock model) with nearest-neighbor spin interactions on a discrete lattice without any externally applied magnetic field is defined as 

\begin{equation}
\mathcal{H} = - J \sum_{< ij >} \sigma_{i}  \cdot \sigma_{j} = - J \sum_{< ij >} \cos \left( \theta_{i} - \theta_{j} \right).
\end{equation}

Here, $J$ represents the coupling constant between interacting spins which is set to unity and thereby favours ferromagnetic order. The sum in Eq.~(1) runs over all neighboring lattice sites $\langle ij \rangle$. We consider square lattices in two dimensions with a total number of $L^2$ spins, where $L$ the lateral extension of the system which we refer to as the ``system size". The spins $\sigma_{i} = \mathbf{e}_{x}\cos (\theta_{i}) +  \mathbf{e}_{y} \sin (\theta_{i}) = [\cos (\theta_{i}), \sin (\theta_{i})]$ are represented as two-component unit vectors in the $x-y$ plane located on discrete and equidistant positions $i$ on the lattice. Within the $q$-state vector Potts model, the possible angles $\theta_{i} \in [0,2 \pi]$ of the spins are given by

\begin{equation}
\theta_{i} = \frac{2 \pi a}{q},
\end{equation}

where the integers $a = 0, 1, 2, ..., q-1$ determine the possible orientations of the spins. At $q = 2$, the model reduces to the classical Ising model with up-down ($Z_2$) spin symmetry, whereas in the limiting case $q \rightarrow \infty$, the model corresponds to the XY model where the spin orientations are continuous within the plane. In what follows, we focus on the cases $q = 4$ and $q \rightarrow \infty$. For, $q = 4$ the model (i.e., the Ashkin-Teller model) shows a second-order phase transition from a PM to a FM phase similar to the Ising model, yet with different characteristics, i.e., different critical exponents \cite{Wu_1982}. In contrast, in the two-dimensional XY model (where $q \rightarrow \infty$), there exists no long-range ordered FM phase at finite temperatures as stated by the Mermin-Wagner theorem \cite{Wagner_1966}. Instead, the sytem undergoes an infinite-order BKT transition from a PM to a BKT phase. 

\begin{figure}
\centering
\includegraphics[width=0.5\linewidth]{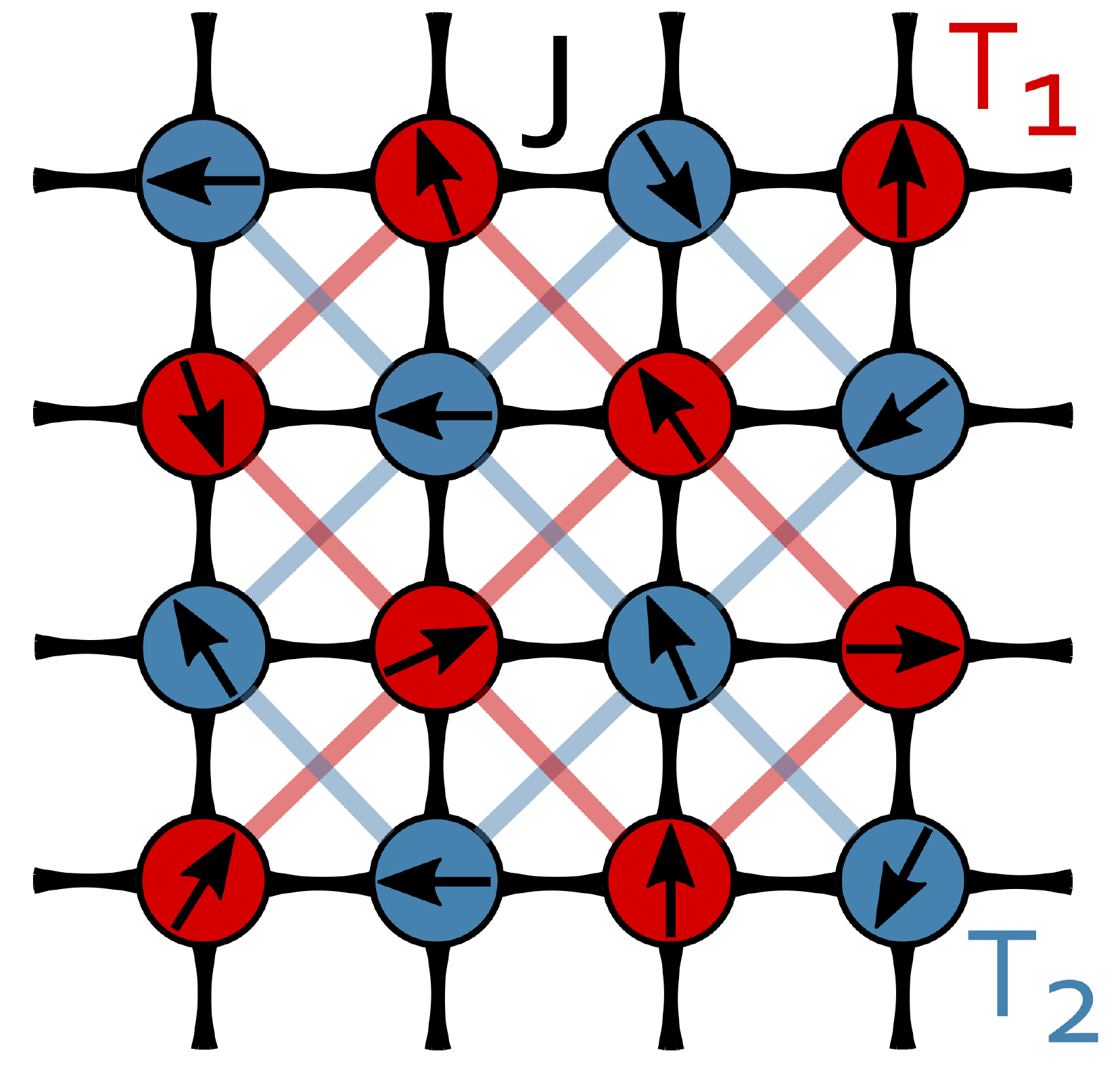}
\caption{Illustration of the $q$-state vector Potts model on a two-dimensional square lattice with periodic boundary conditions. The spins are coupled to heat baths at two different temperatures $T_{1}$ and $T_{2}$, here indicated by the colors red and blue, respectively, with a checkerboard arrangement. Black lines represent the nearest-neighbor interactions of strength $J$, while red and blue lines highlight the two sublattices formed by the coupling to heat baths of different temperatures.}
\end{figure}

In order to study the dynamical evolution of the system in presence of thermal noise, we perform Monte-Carlo simulations with single spin-flip Glauber dynamics. Here, the rate $w_{\mu \nu}(i)$ for a transition of a randomly chosen spin $\sigma_i$ from state $\nu$ (before the spin flip) to $\mu$ (after the spin flip) depends only on the temperature $T_k$ of the heat bath the considered spin is coupled to, as well as on the energy difference $\Delta \mathcal{H} = \mathcal{H}(\nu) - \mathcal{H}(\mu)$ related to a flip of spin $\sigma_i$, that is,

\begin{equation}
w_{\mu \nu}^{i} = \frac{1}{2} \left[ 1 - \sigma_{i} \tanh \left( \Delta \mathcal{H} / T_k  \right) \right].
\end{equation}

If all spins $\sigma_{i}$ are exposed to a single heat bath at temperature $T$, the system (that is initially prepared in a configuration with random spin orientations) eventually reaches a state of thermal equilibrium and thus, does not produce entropy, $\Pi = 0$ (see Sec.~\ref{sec:EP} for a definition of entropy production). In contrast, here we drive the system into a nonequilibrium steady state by coupling the spins $\sigma_{i}$ to two different heat baths $T_k$ ($k = 1,2$) which are kept at temperatures $T_{1}$ and $T_{2}$. With this setup, the system is out of equilibrium whenever $T_{1} \neq T_{2}$. There is a constant heat flux $\dot{Q}$ from the hotter to the colder heat reservoir that goes along with a constant rate of entropy production, $\Pi > 0$ (see Sec.~\ref{sec:EP}). Our setup splits the system into two sublattices, $\mathcal{L}_{1}$ and $\mathcal{L}_{2}$, each containing all spins connected to the bath at $T_{1}$ or $T_2$, respectively. All four nearest-neighbors of a spin $\sigma_{i}$ are coupled to the respective other heat bath, yielding a checkerboard configuration as illustrated in Fig. 1. In the following, we fix $T_{2}$, but vary $T_{1}$ and calculate all observables as function of the mean temperature $T = (T_{1} + T_{2}) / 2$.

\section{Entropy production in the vector Potts model}\label{sec:EP}
A key quantity that distinguishes systems out of thermal equilibrium from those in equilibrium is the constant net production of entropy. In general, the dynamical evolution of physical systems that possess a finite set $\nu \in \Omega$ of discrete microstates (e.g., the spin configurations of the vector Potts model at finite $q$) can be described as continuous time Markov Chains. For such systems, the time-dependent system (or Shannon) entropy \cite{Shannon_1948} is given by the Boltzmann-Gibbs expression

\begin{equation}
S(t) = - \sum_{\nu} p_{\nu}(t) \ln  \left[ p_{\nu}(t) \right].
\end{equation}

Here, the sum runs over all microstates (i.e., spin configurations) and $p_{\nu}(t)$ represents the occupation probability of state $\nu$ at time $t$. By coupling the system to an infinitely large heat bath at temperature $T$ (that always maintains a state of thermal equilibrium), we can formulate an expression for the time-dependent change of entropy \cite{Seifert_2012} that one the one hand, originates from the total system internal production of entropy $\Pi(t)$ and, on the other hand, by the exchange of entropy $\Phi(t)$ with the environment, 

\begin{equation}
\partial_{t} S(t) = \Pi(t) - \Phi(t).
\end{equation}

In order to formulate explicit expressions for $\Pi(t)$ and $\Phi(t)$, we make use of the fact that the (generally time-dependent) occupation probabilities $p_{\nu}(t)$ obey a master equation 

\begin{equation}
\partial_{t} p_{\nu}(t) = \sum_{\mu} \left[ w_{\nu \mu}(t) p_{\mu}(t) -  w_{\mu \nu}(t) p_{\nu}(t) \right].
\end{equation}

The change $\partial_{t} p_{\nu}(t)$ stems first, from the total incoming probability flow $\sum_{\mu \neq \nu} w_{\nu \mu}(t) p_{\mu}(t)$ consisting of all possible state transitions $\mu \rightarrow \nu$ happening with transition rates $w_{\nu \mu}(t)$. The second contributive to $\partial_{t} p_{\nu}(t)$ is the total outgoing probability flow  $\sum_{\mu} w_{\mu \nu}(t) p_{\nu}(t)$ due to transitions $\nu \rightarrow \mu$. 

For systems in thermal equilibrium, the detailed balance (DB) condition, $w_{\nu \mu} p_{\mu} = w_{\mu \nu} p_{\nu}$, holds for all $\mu$ and $\nu$. When DB is violated, there are non-vanishing local probability flows between certain microstates, i.e., $w_{\nu \mu} p_{\mu} - w_{\mu \nu} p_{\nu} \neq 0$. As a consequence, the system constantly produces entropy $\Pi(t) > 0$, which is given by \cite{Schnakenberg_1976}

\begin{equation}
\Pi(t) = \frac{1}{2} \sum_{\mu, \nu} \left[w_{\nu \mu}(t) p_{\mu}(t) - w_{\mu \nu}(t) p_{\nu}(t)  \right] \ln \frac{w_{\nu \mu}(t) p_{\mu}(t)}{ w_{\mu \nu}(t) p_{\nu}(t)}.
\end{equation}
Equation~(7) obeys the thermodynamically expected properties: $\Pi(t)$ nullifies in thermal equilibrium, and is strictly positive otherwise, in accordance with the second law of thermodynamics. For nonequilibrium stationary states (no time-dependency), $\Pi(t) = \Pi$, one has $\partial_t S = 0$ and consequently from Eq.~(5), $\Pi = \Phi$. Additionally, Eq.~(7) reduces to

\begin{equation}
\Phi =  \Pi =  \sum_{\mu, \nu} w_{\nu \mu} p_{\mu} \ln \frac{w_{\nu \mu}}{w_{\mu \nu}} =  \sum_{\mu, \nu} j_{\nu \mu} \ln \frac{w_{\nu \mu}}{w_{\mu \nu}}.
\end{equation}

Equation (8) can be computed numerically by averaging over many transitions $\mu \rightarrow \nu$ from the current state $\mu$ of the Markov Chain (i.e., the current spin configuration of the lattice) in the steady state. In spin systems with discrete spin orientations like the vector Potts model with finite $q$, state transitions $\mu \rightarrow \nu$ correspond to the flipping of a randomly chosen spin $\sigma_i$ on lattice site $i$. Thus, the sum in Eq.~(8) can be written as an average over all lattice sites

\begin{equation}
\Phi = \Pi = \sum_{i} \sum_{\nu} \biggl< w_{\nu \mu}^{i} \ln \frac{w_{\nu \mu }^{i}}{w_{\mu \nu}^{i}} \biggl>.
\end{equation}
Here, $w_{\nu \mu}^{i}$ [see Eq.~(3)] corresponds to the Glauber-type flipping rate of the spin $\sigma_i$ on lattice site $i$ (which is connected to the heat bath at $T_k$), inducing a transition from the current state $\mu$ to state $\nu$ due to a change of the current orientation $\theta_i$ of spin $\sigma_i$ to any other allowed one. The steady exchange of entropy with the environment results from the net heat flux $\dot{Q}$ from the hotter to the colder sublattice. We here employ the sign convention 
	$\dot{Q}>0$ for the heat flow from hot to cold.
	Due to energy conservation (and because no external fields, forces or further gradients act on the system), all of the three relevant heat flows transport the same amount of energy per timestep: the flow from the hotter heat bath $T_1$ (here we assume for a moment $T_1>T_2$) to the corresponding sublattice $\mathcal{L}_1$, the flow from  $\mathcal{L}_1$ to $\mathcal{L}_2$, and the heat flow from  $\mathcal{L}_2$ to the cold bath at $T_2$ (or everything reversed, if $T_2>T_1$). This amounts to an overall entropy flow to the environment of $\Phi=|(\dot{Q}/T_2) - (\dot{Q}/T_1)|=\dot{Q}\,|T_2-T_1|/(T_1T_2)$.

Equation~(9) can be used to calculate the mean of the entropy production rate, $\Pi$, for systems with a finite set $\Omega$ of discrete microstates (i.e., the vector Potts model with $q = 4$) by means of Monte-Carlo simulations. In Sec.~\ref{sec:results}, we show results for the entropy production rate per spin, which is given by $\pi = \Pi / L^{2}$.

A problem with Eq.~(9) is that it can not be used to calculate $\Pi$ in systems with continuous degrees of freedom, as it is the case for the   vector Potts model with $q \rightarrow \infty$ (XY model). As an alternative, we calculate the entropy production rate by following individual stochastic trajectories consisting of consecutively executed state transitions $s_{n-1} \to s_{n}$ between microstates as the system dynamically evolves via the reorientation of single spins $\sigma_i$ \cite{Seifert_2012}. Such trajectories correspond to a sequence ($s_0 \rightarrow s_1 \rightarrow s_2 ... \rightarrow s_{n-1} \rightarrow s_n ... \rightarrow s_{l-1} \rightarrow s_l$) consisting of $l$ state transitions, each connected with a transition rate, $w_{s_{n} s_{n-1}}^{i}$, given by Eq.~(3). Note that each individual state that is part of the trajectory simply corresponds to one of the microstates of the system, $s_n \in \Omega$, i.e., to one of the possible spin configurations. Since there are infinitely many states for $q \rightarrow \infty$, we make use of the fact that each transition $s_{n-1} \rightarrow s_n$ [for which we know the transition rate $w_{s_{n} s_{n-1}}^{i}$ according to Eq.~(3)] along the stochastic path due to a random reorientation of a spin $\sigma_i$ is associated with a (stochastic) entropy exchange with the environment $\Delta \phi = \ln \left( w_{s_{n} s_{n-1}}^{i} / w_{s_{n-1} s_{n}}^{i} \right)$ \cite{Seifert_2012} and an associated local heat exchange of $\Delta \phi /T_k $. As a consequence, the total change of entropy along an individual stochastic path consisting of $l$ transitions reads

\begin{equation}
\Delta \phi(l) = \sum_{n = 1}^{l} \ln \frac{ w_{s_{n} s_{n-1}}^{i}}{ w_{s_{n-1} s_{n}}^{i}}.
\end{equation}

In the limit of infinitely long trajectories, $l \rightarrow \infty$, Eq.~(10) divided by the length $l$ of the trajectory becomes identical to the ensemble averaged medium entropy production rate $\Phi$ due to the ergodicity of the system. In a steady state this is further identical to the total entropy production rate as calculated according to Eq.~(9)

\begin{equation}
\lim_{l \rightarrow \infty} \frac{\Delta \phi(l)}{l} = \Phi = \Pi.
\end{equation}

In addition, we also use Eq.~(10) to obtain distributions $P[\Delta \phi(l=100)]$ for the $4$-state model as well as the version with $q \rightarrow \infty$ (see Sec.~\ref{sec:details} for details).

\section{Measurement details and parameter settings}\label{sec:details}

In the present study, simulations of the vector Potts model with nearest-neighbor interactions are performed on square lattices with lateral extension ranging from $L = 16$ to $L = 96$. This means that we consider $L^2= 256$ up to $L^2 = 9216$ spins. Before calculating any physical quantity, we first let the system evolve for $5 \times 10^{4}$ Monte Carlo steps (MCS) [where one MCS consists of $L^2$ spin flip attempts with spin-flip rates $w_{\mu \nu}^{i}$ according to Eq.~(3)] to assure that the system has reached a steady state. We then let each system further evolve up to a maximum of $10^{6}$ MCS and use (depending on the system size $L$) between $100$ and $1000$ realizations for each parameter setting (i.e., combination of $T_{1}$ and $T_{2}$) in order to guarantee the convergence of average quantities.

In order to quantify the phase behavior of the system, we calculate the magnetic order parameter

\begin{equation}
m = \frac{1}{L^{2}} \sqrt{\left( \sum_{i} \cos \theta_{i} \right)^{2} + \left( \sum_{i} \sin \theta_{i} \right)^{2}}.
\end{equation}

The value of $m \in [0,1]$ is a measure for the spin ordering in the system. For perfect order, $m = 1$, while in a completely disordered system, $m = 0$. We also define the magnetic order parameters for the two sublattices $\mathcal{L}_{1}$ and $\mathcal{L}_{2}$

\begin{equation}
m_k = \frac{1}{2 L^{2}} \sqrt{\left( \sum_{i \in \mathcal{L}_{k}} \cos \theta_{i} \right)^{2} + \left( \sum_{i \in \mathcal{L}_{k}} \sin \theta_{i} \right)^{2}},
\end{equation}

where the index $k = 1,2$ denotes the respective sublattice.

To precisely determine the value of the critical temperature $T_{c}$ where the phase transition (from the PM to FM or to the BKT phase) sets in, we compute the fourth-order Binder cumulant \cite{Binder_1981} of the magnetic order parameter $m$ 

\begin{equation}
U_{4} = 1 - \frac{\langle m^{4} \rangle}{3 \langle m^{2} \rangle^{2}},
\end{equation}

which is universal at criticality. The critical value $T_{c}$ of the control parameter is given by the intersection point of $U_{4}$ for different lateral sizes $L$ of the system.

In addition, we calculate the specific heat per lattice site which is given by

\begin{equation}
C_{v} = \frac{1}{  T^{2}} \left[ \langle E^{2} \rangle - \langle E \rangle^{2} \right],
\end{equation}
what can also be expresses as $\mathrm{d}\langle E\rangle/\mathrm{d}T$. Here, $\langle E \rangle$ corresponds to the average energy per spin. The specific heat is known to exhibit power-law scaling as the critical critical temperature $T_{c}$ is approached from the PM disordered phase. Additionally, $C_v$ peaks at $T_{c}$ where its maximum shows power-law scaling as function of $L$ which is often universal \cite{Mccoy_2004}.

The average EP rate per spin, $\pi = \Pi / L^{2}$, is calculated according to Eq.~(10) for the $4$-state vector Potts model, while Eq.~(11) is used in the case of the XY model ($q \rightarrow \infty$) where the spin orientation is continuous. However, $\pi$ can also be obtained from Eq.~(11) for the $4$-state model. In both cases, we check whether the change of $\pi$, with respect to the control parameter $T$, ${\mathbf{d} \pi}/{\mathbf{d} T}$, shows universal features (similar to the specific heat $C_{v}$) regarding its scaling behavior as function of system size $L$ around the critical point $T_{c}$ of the phase transition. 

Distributions $P(\phi)$ of the change of entropy $\phi = \Delta \phi(l)$ for trajectories of length $l = 100$ are obtained via Eq.~(10) for $q = 4$ and $q \rightarrow \infty$. We calculate $\phi$ for the whole lattice and the individual sublattices $\mathcal{L}_1$ and $\mathcal{L}_2$, respectively. This is done by defining trajectories that only account for the change of entropy induced by state transitions due to a reorientation of spins which are connected to the respective heat bath $T_k$ ($k$ = 1,2). The distributions $P(\phi)$ are obtained from at least $10^{7}$ trajectories.

\section{Results}\label{sec:results}

In this section, we present a numerical investigation of the nonequilibrium vector Potts model with discrete ($q = 4$) and continuous ($q \rightarrow \infty$) symmetry. In both cases, we find that the nonequilibrium model exhibits the same type of phase transition as in the equilibrium case. We therefore focus for $q = 4$ on the transition from the spin-disordered PM to the spin-ordered FM phase, whereas for $q \rightarrow \infty$ we analyze the BKT-like transition from the disordered PM to the quasi long-range ordered BKT phase. In both cases the transitions are continuous in the order parameter. To characterize the critical behavior, we study the specific heat and the entropy production, and we compare the results in the vicinity of the critical point for both kinds of phase transition.

\subsection{Phase transition in the discrete vector Potts model with $q = 4$}
Before we numerically investigate the phase transition of our nonequilibrium model, let us briefly review some important properties of the equilibrium version of the $4$-state vector Potts model. It is well known \cite{Wu_1974,Domany_1981,DeFellcio_1982} that the equilibrium model exhibits a second-order phase transition at $T_c^{eq} = 1.13$, which is half the critical temperature of the classical Ising model ($T_c^{eq} = 2.26$) that exhibits up-down symmetry. The critical exponents of the $4$-state version are different from the Ising model \cite{Wu_1982}. 

To begin with the analysis of the nonequilibrium model, we consider the behavior of the ensemble-averaged magnetization $m$ [see Eq.~(12)], which serves as a global order parameter. Figure~2 displays $m$ as function of the mean temperature $T$ for four different (fixed) values of $T_{2}$ and various system sizes ranging from $L = 16$ to $L = 96$. The most prominent observation is that while decreasing the mean temperature from high values, the order parameter increases and eventually approaches its maximum value, $m = 1$ (reflecting perfect spin order). This implies the existence of a stable FM phase at low bath temperatures, although the system is clearly out of equilibrium.

\begin{figure}
	\centering
	\includegraphics[width=0.99\linewidth]{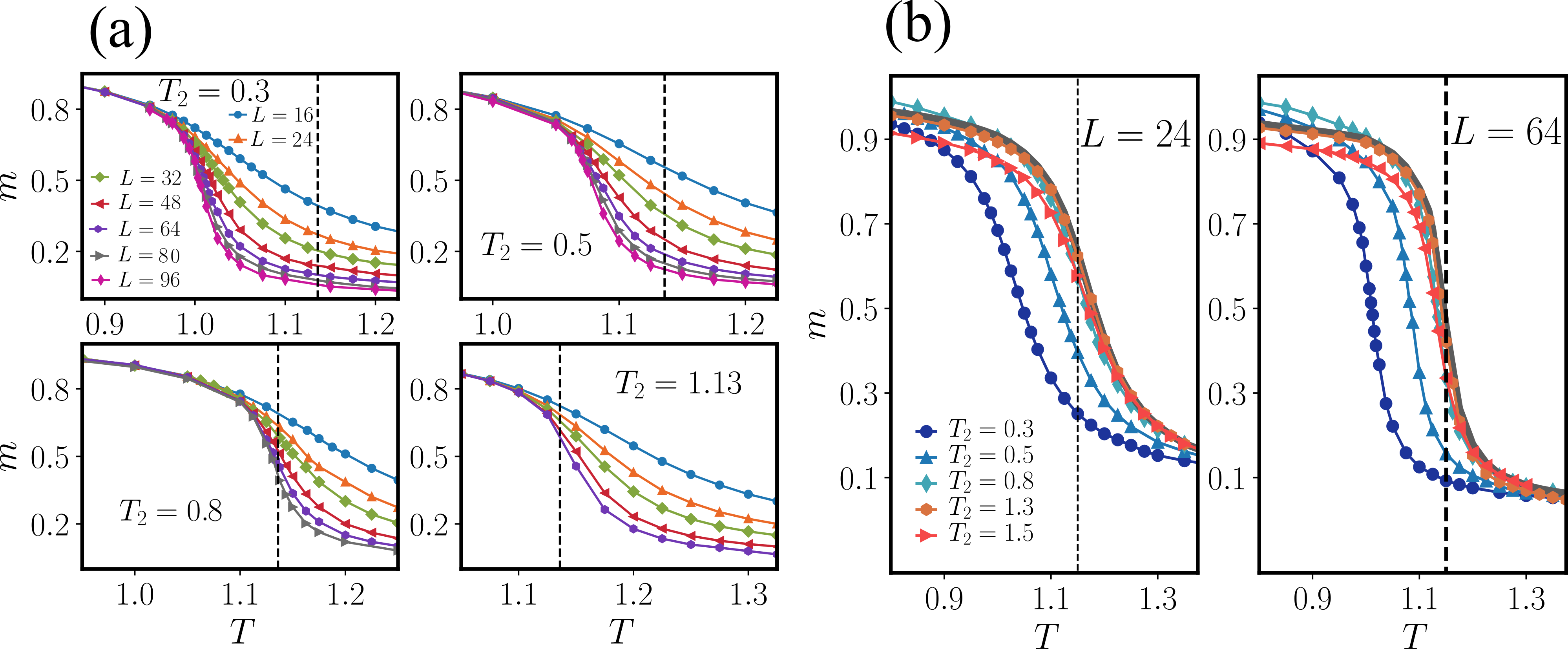}
	\caption{$(a)$ The ensemble-averaged magnetization $m$  vs. mean temperature $T=(T_1+T_2)/2$ in the vector Potts model with $q = 4$, for system sizes ranging from $L = 16$ to $L = 96$ (different colors and symbols) and fixed value for the temperature $T_2$ in each panel. $(b)$ The magnetization $m$ in the model with $q = 4$ as function of $T$ for $L = 24$ and $L = 64$ and different values of $T_{2}$ ranging from $T_{2} = 0.3$ to $T_{2} = 1.5$. The solid gray lines correspond to $m$ in the equilibrium model where $T = T_1 = T_2$. The dashed vertical lines in both panels indicate the critical temperature $T_{c}^{eq} = 1.13$ of the equilibrium model.
	}
\end{figure}

Figure~2(a) indicates that the nonequilibrium phase transition occurs at a temperature which is comparable to the one of the equilibirum model, $T_c^{eq} = 1.13$. However, a closer inspection reveals that the precise value of $T_{c}$ depends on the fixed temperature $T_{2}$ in such a way that $T_{c}$ becomes smaller as $T_{2}$ is shifted away from the critical value $T_{c}^{eq}$ of the equilibrium model. This conspicuousness is further confirmed by Fig.~2(b), where (for $L = 24$ and $L = 64$) $m$ is plotted as function of $T$ for different values of $T_{2}$. Remarkably, the shift of the temperature region (compared to the equilibrium model) where $m$ as function of $T$ increases to large  values (indicating the emergence of spin order) is found to be equally large for both system sizes $L$. This, in turn, signals that the temperature shift of the curves is not a finite size effect (which would vanish for $L \rightarrow \infty$), but an actual property of this nonequilibrium vector Potts model. A further interesting observation from Fig.~2(b) is that the nonequilibrium vector Potts model studied here displays an ordered phase, even if one of the heat bath's temperatures is higher than the critical temperature $T_c^{eq}$ of the corresponding equilibrium model. However, the critical mean temperature is always below the equilibrium value, as we will discuss below. 

To precisely analyze the dependency of $T_{c}$ on $T_2$ (and $T_1$), we compute the Binder cumulant $U_{4}$ as function of $T$ for different values of $L$ [see Eq.~(14) and below]. Figure~3(a) shows the crossing of the respective lines for two exemplary temperatures $T_2$, clearly confirming the aforementioned shift of $T_c$.

\begin{figure}
\centering
	\includegraphics[width=0.99\linewidth]{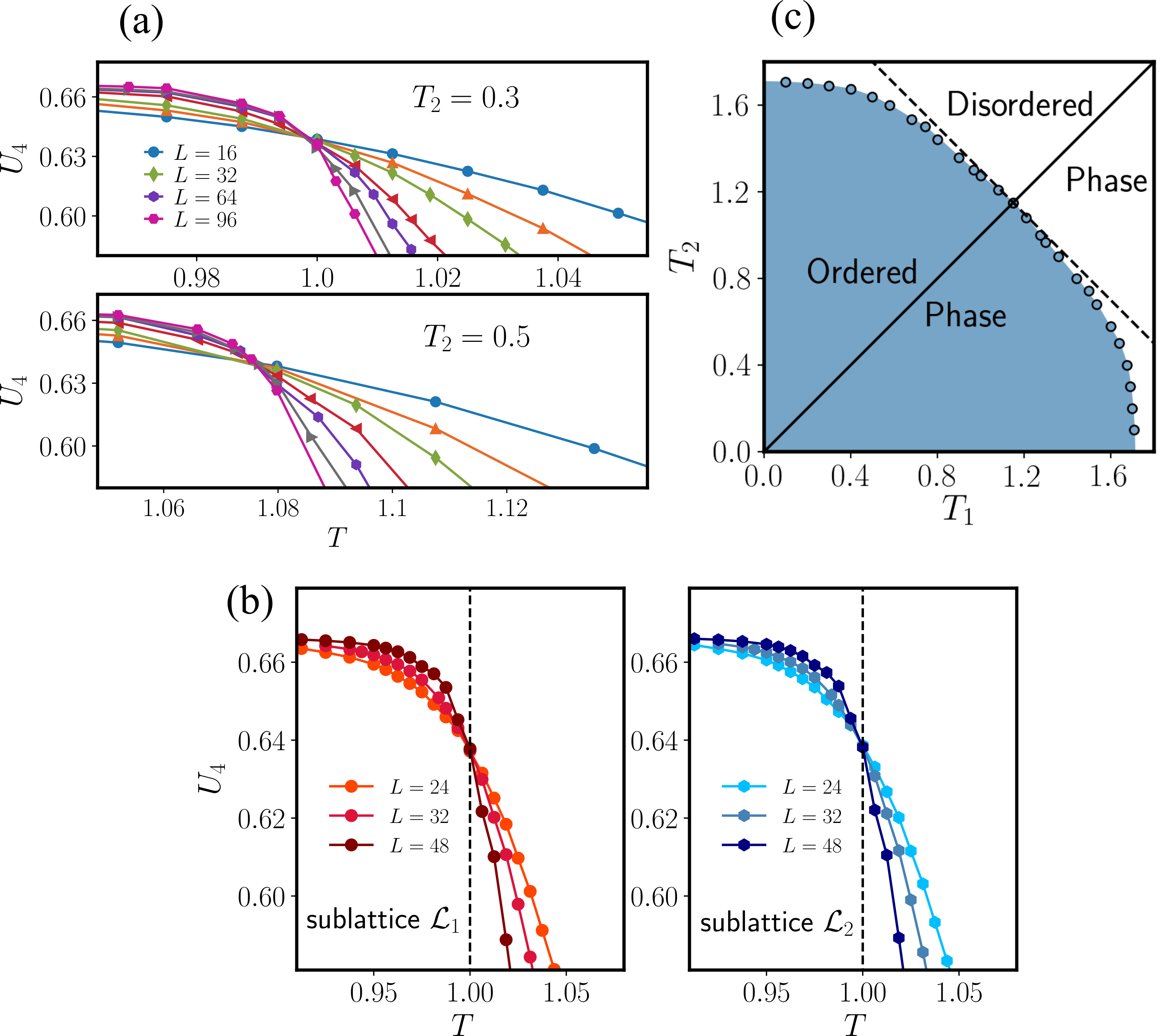}
	\caption{(a) Binder cumulant $U_{4}$ in the vector Potts model with $q = 4$ and system sizes from $L = 16$ to $L = 96$. The intersection point marks the value of the critical temperature $T_c$. At $T_2 = 0.3$ we find $T_c =0.997$$(4)$, while at $T_2 = 0.5$ the lines intersect at $T_c=1.075(8)$. (b) Binder cumulant of the two sublattices $\mathcal{L}_1$ and $\mathcal{L}_2$ with $q = 4$ for $T_{2} = 0.3$ and $T_1$ ranging from $T_1 = 1.5$ to $T_1 = 1.9$ for $L = 24$ to $L = 48$. The dashed lines mark the critical value $T_c$ where the Binder cumulant intersects. (c) Phase diagram of the nonequilibrium vector Potts model with $q = 4$ showing the boundary (circles) between the FM ordered (indicated by the blue shaded region) and the PM disordered phase as function of $T_{1}$ and $T_{2}$. The critical temperatures have been obtained from the crossing of the Binder cumulant for different system sizes $L$.}
\end{figure}

Since the spins $\sigma_i$ are coupled to different heat baths, one might expect differences in the phase behavior of the two sublattices $\mathcal{L}_1$ and $\mathcal{L}_2$. However, as one can see in Fig.~3(b), the Binder cumulant intersects at the same temperature $T_c$ in both sublattices. This shows that the transition from the paramagnetic to the ferromagnetic phase occurs collectively in the entire system at the same temperature $T_c$.

For an overview of the critical temperatures in the plane spanned by $T_1$ and $T_2$, we now look at the nonequilibrium phase diagram plotted in Fig.~3(c). The diagonal (black solid) line where $T_1 = T_2$ corresponds to the equilibrium model. For the nonequilibrium system ($T_1 \neq T_2$), $T_c$ depends on $T_1$ and $T_2$ approximately linearly in the vicinity of equilibrium ($T_1 \approx T_2$) but the dependency becomes strongly nonlinear when $T_1 \gg T_2$ or $T_2 \gg T_1$. This is clearly seen when one compares the actual phase boundary with the dashed curve corresponding to the line along which $T = T_c^{eq}$ holds, i.e., $T_2 = 2T_c^{eq}-T_1$. 
One can further see that when $T_1=T_2$ (i.e., in the equilibrium model), the phase transition occurs at the highest mean temperature $T_c^{eq}$. As soon as there is a temperature difference between the two sublattices, the nonequilibrium phase transitions occur at a lower critical temperature, which deviates from $T_c^{eq}$ the more as the difference $\Delta T = |T_1 - T_2|$ between $T_1$ and $T_2$ increases. Moreover, there exists a new type of critical temperature $T_c^{*} = 1.700(2)$ with the following property: If one sublattice has a temperature higher than $T_c^{*}$, global order is destroyed, irrespective of the temperature of the other sublattice. 

Physically, one may understand the phase behavior in the following manner. When heating the system up in the presence of a temperature difference $\Delta T$ between the sublattices, disorder is favored already at lower system-averaged temperatures $T$, showing that a smaller amount of thermal noise destroys the long-range order. Consistent with our physical intuition, a breaking of the translational symmetry (by the temperature difference between the sublattices) reduces the stability of long-range order. Note that this is in sharp contrast to the situation where an homogeneous external magnetic field acts on the system (breaking the up-down symmetry) which increases the stability of long-range ordering. 

\subsubsection{Critical behavior of specific heat}

After the determination of $T_{c}$ and its dependency on the temperatures of the two heat baths, we now turn to the investigation of the thermodynamic properties of our nonequilibrium spin model in the vicinity of the phase transition. First, we calculate the specific heat $C_{v}$ [see Eq.~(16)] as function of the mean temperature $T$ for different values of $L$ and $T_{2}$. This quantity is commonly considered in order to characterize second-order phase transitions. As can be seen in Fig.~4(a), $C_{v}$ peaks at a temperature very close to the values of $T_c$ that we have previously determined via the Binder cumulant (recall Fig.~3).

\begin{figure}
\centering
	\includegraphics[width=0.99\linewidth]{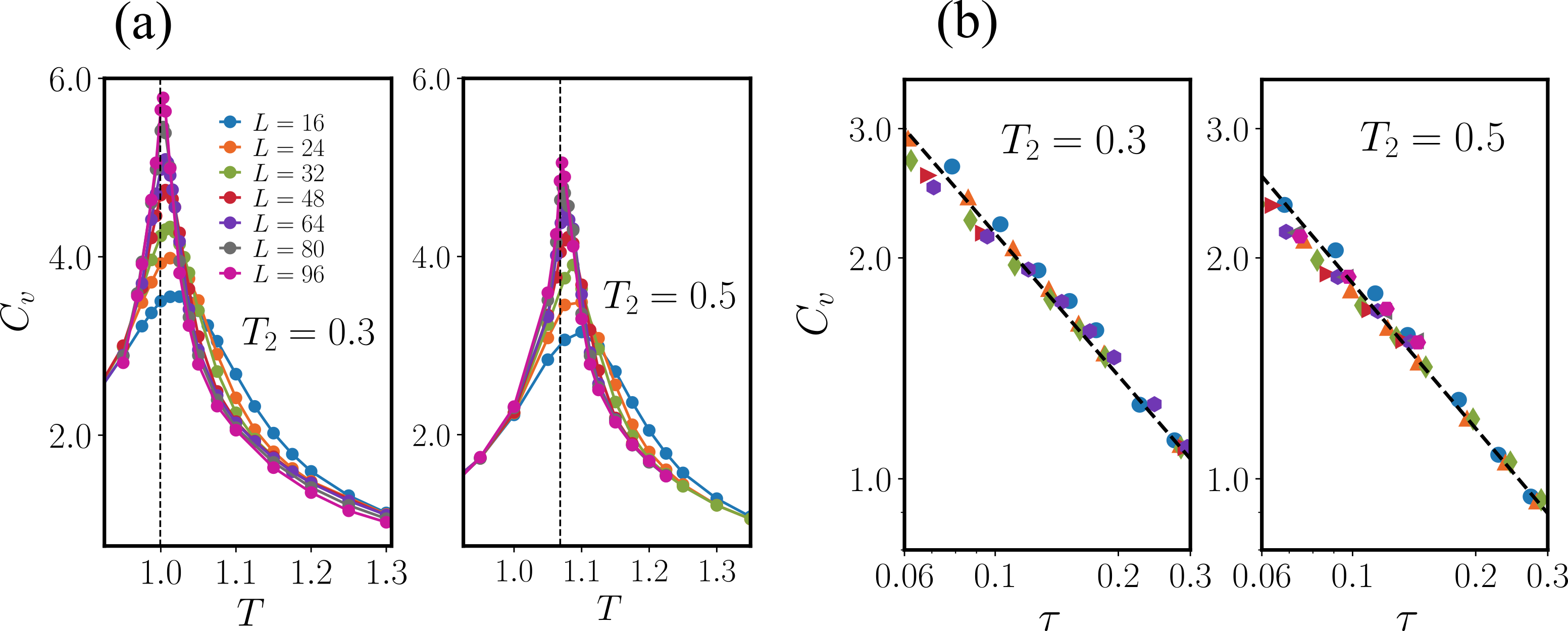}
	\caption{$(a)$ The specific heat $C_{v}$ vs. $T$ for $T_{2} = 0.3$ and $T_2 = 0.5$ and system sizes ranging from $L = 16$ up to $L = 96$. $(b)$ Power-law scaling of the specific heat $C_{v}$ vs. the reduced temperature $\tau = |1 - T/ T_c|$ in the disordered phase for fixed $T_{2}$ and system sizes ranging from $L = 16$ up to $L = 96$ indicated by different colors and symbols. For both values of $T_2$ the dashed black line follows $\sim 2/3$.}
\end{figure}

For both depicted values of $T_{2}$, the precise location of the peak depends on $L$ in such a way that as the system size is increased, the temperature where the peak is located decreases and approaches $T_{c}$. We suspect that the peak is exactly at $T_c$ in the limit $L\to \infty$, as it is well-known for the equilibrium version of this model.
In thermal equilibrium, the specific heat of the model is further known to show universal scaling behavior with respect to the temperature, i.e., 
\begin{equation}
C_\nu \sim |1-T/T_c|^{-\alpha}
\end{equation}
with $\alpha=2/3$ in the disordered phase \cite{Wu_1982}. Interestingly, we find that the nonequilibrium model also displays a power-law divergence of $C_v$. Moreover, the critical exponent is the same as in equilibrium, irrespective of the value of the temperature gradient $\Delta T = |T_2 - T_1|$ among the two heat baths (as long as $T_1$ and $T_2 \le T_c^{*}$). This is exemplarily illustrated for $T_2 = 0.3$ and $T_2 = 0.5$ in Fig. 4(b), where $C_v$ is plotted for different system sizes (from $L = 16$ to $L = 96$) as function of the reduced temperature $\tau = |1 - T/T_c|$ together with straight (black dashed) lines (with slope $- 2/3$). We checked the scaling behavior for various additional values of $T_2$ and all of them show a power-law scaling with $\alpha = 2/3$, demonstrating the robustness of the critical exponent under nonequilibrium conditions. To analyze the critical behavior based on our numerical data in detail, we employ the finite-size scaling technique. To this end, we consider the positions of the peaks of $C_\nu$, which give an approximation for the critical temperature as function of the system size $L$. For the equilibrium $4$-state vector Potts model on a square lattice, this quantity scales as $\sim L^{-\nu}$, with the corresponding critical exponent $\nu=2/3$~\cite{Tobochnik_1982}. Also for the nonequilibrium model we obtain $\nu = 2/3$ for all values of $T_2$, consistent with the well-known scaling law $\nu d = 2-\alpha$ \cite{Kadanoff_1966} (where $d = 2$ is the spatial dimension of the lattice).

\subsubsection{Critical behavior of total entropy production}

Let us now consider the behavior of the total EP which is a direct measure for irreversibility in the sense that it quantifies the distance from equilibrium. To start with, we find that the total EP per spin is always positive, $\pi > 0$, whenever $T_{1} \neq T_{2}$. Moreover, $\pi$ is a convex function of the mean temperature $T$ with minimum at the equilibrium point $T=T_{1}=T_{2}$, where $\pi = 0$ [consistent with Eq.~(7)]. This can be seen in Fig.~5(a), which depicts $\pi$ vs. $T$ for an exemplary setting of $T_{2} = 1.5$ and $L = 32$ around the equilibrium mean temperature $T = T_1 = T_2 = 1.5$. 

\begin{figure}
\centering
	\includegraphics[width=0.8\linewidth]{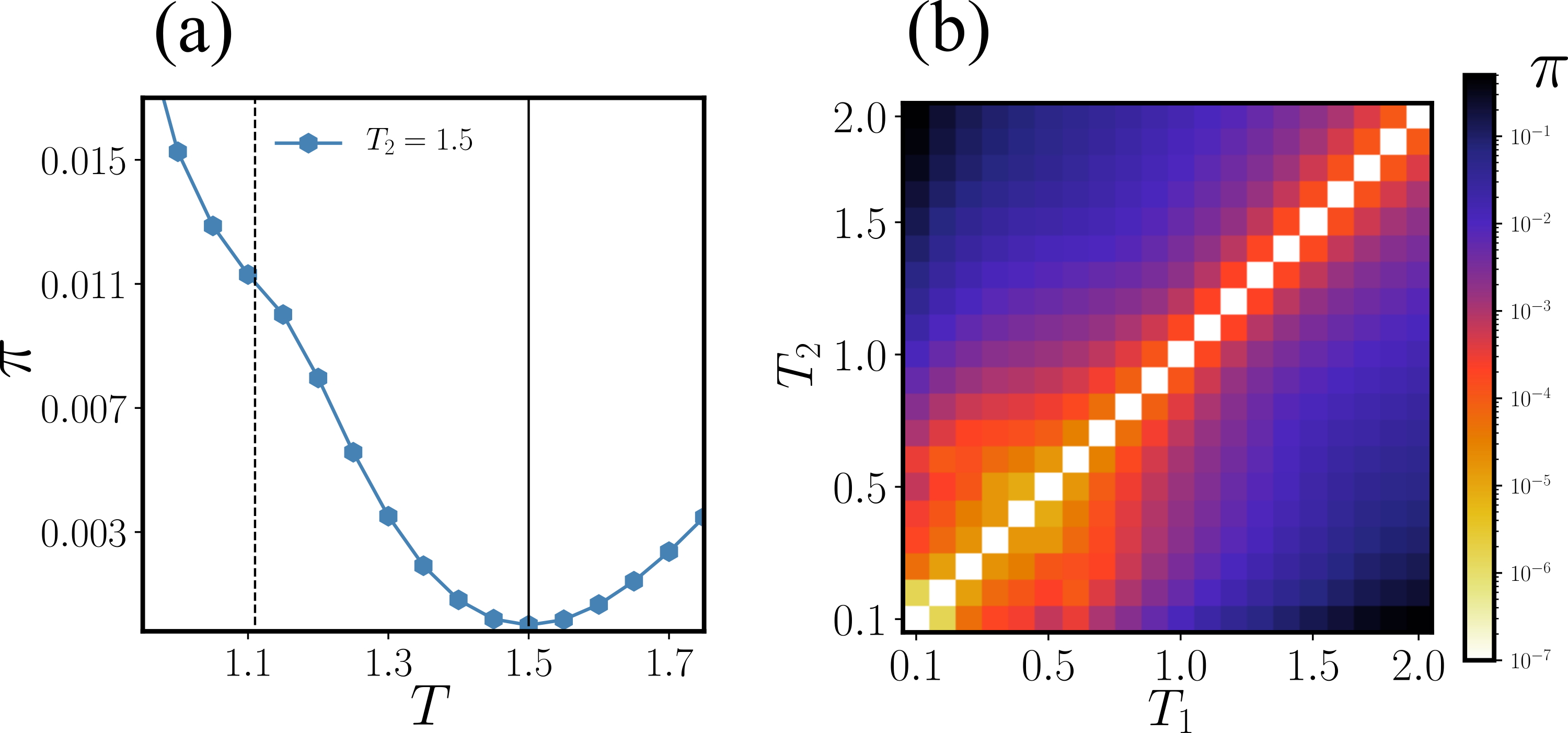}
	\caption{$(a)$ The EP rate per spin, $\pi$, as function of the mean temperature $T$ for fixed $T_{2} = 1.5$ and system size $L = 32$ in the model with $q = 4$. The solid black line corresponds to the equilibrium point where $T_1 = T_2$ and the dashed black line marks the critical temperature $T_c$ of the FM to PM phase transition. $(b)$ Heatmap of the EP rate per spin, $\pi$, in the vector Potts model with $q = 4$ on a lattice of size $L = 32$ for temperatures of the two sublattices ranging from $T_1 = T_2 = 0.1$ up to $T_1 = T_2 = 2.0$.}
\end{figure}

Depending on whether $T_2$ is higher or lower than $T_c^{eq}$, the phase transition of the nonequilibrium model lies below or above that minimum (which is always located at $T = T_1 = T_2$). In other words, if $T_2>T_c$, which is the situation considered in Fig.~5(a), the nonequilibrium phase transition occurs at the left hand side of the minimum, whereas if $T_2<T_c$, the transition occurs at the right hand side of it. Interestingly, we observe that $\pi$ as function of $T$ shows a bump around $T_c= 1.11$. In that sense, the function $\pi(T)$ itself already signals the occurence of the phase transition.

The dependency of entropy production rate per spin on the temperatures of the two heat baths is plotted in Fig.~5(b), which shows $\pi$  for different combinations of $T_1$ and $T_2$ ranging from $0.1$ up to $2.0$. As expected, $\pi = 0$ whenever $T_1 = T_2$ [i.e., no temperature gradient $\Delta T$ is present and detailed balance is fulfilled, see Eq.~(7)]. In contrast to this, there is always a positive rate of entropy production ($\pi > 0$) when $T_1 \neq T_2$, consistent with the special case considered in Fig.~5(a). As the gradient $\Delta T$ increases, the entropy production rate increases roughly $\pi \sim \Delta T^{2}$, no matter whether the system is in the PM or the FM phase. 

\begin{figure}
	\includegraphics[width=0.99\linewidth]{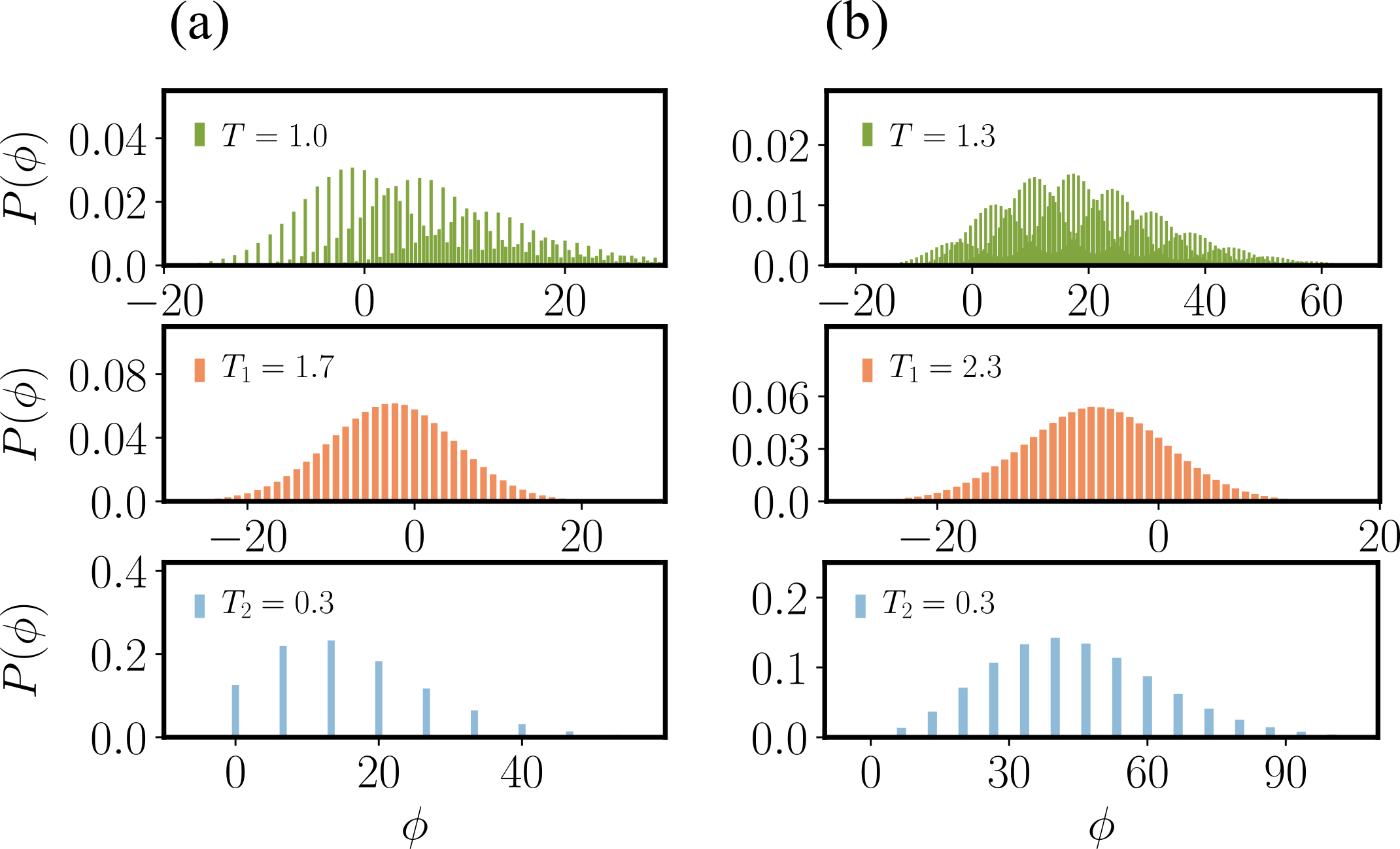}
	\caption{Distribution $P(\phi)$ of the stochastic medium entropy $\phi = \Delta \phi(l)$ that is produced in the system along individual stochastic trajectories of length $l = 100$ in the $4$-state vector Potts model. The panels of (a) show $P(\phi)$ in the vicinity of the phase transition [which is at $T=T_c = 0.997(4)$] for a system with $L = 64$, at $T_1 = 1.7$ and $T_2 = 0.3$. The top panel shows the distribution of the whole system. The middle panel shows the distribution of the medium entropy production of the spins which belong to the sublattice $\mathcal{L}_1$, while the one at the bottom shows the distribution for $\mathcal{L}_2$. (b) shows the corresponding distributions in the PM disordered phase, at $T=1.3>T_{c}$, specifically at $T_1 = 2.3$ and $T_2 = 0.3$.}
\end{figure}

To resolve the EP along individual stochastic trajectories, we plot distributions of the medium entropy production $P(\phi)$. Figure~6 displays numerical results for
$P(\phi)$ at the critical temperature $T_c$ [Fig.~6(a)] and above $T_c$ [Fig.~6(b)], for trajectories of length $l = 100$ (see Sec.~\ref{sec:details} for details).
We consider both, the distribution of the entire system as a whole (top panels), and the separate distributions obtained by restricting our observation to one of the two sublattices, $\mathcal{L}_1$ or $\mathcal{L}_2$, only (middle and bottom panels, respectively). For example, the middle panels show the histograms of all detected values of the medium EP from spins that belong to the sublattice $\mathcal{L}_1$. Overall, the main characteristics seem to be quite similar for the system at and above the phase transition [compare (a) and (b)].
Let us now take a closer look at the different distributions. 
Remarkably, in both cases, $P(\phi)$ for the whole lattice exhibits a multi-peaked structure [see top panels in Fig.~6]. When inspecting the corresponding sublattice distributions, we notice that the multi-peak structure of the whole system appears to arise as a combination of both sublattices. This is reasonable, as the stochastic trajectories of the whole system expectantly include both, many contributions from the hotter sublattice (which flips more often), and some seldom contributions from the colder sublattice. In fact, the multi-peaked structure looks like a convolution of the distributions from the belonging sublattices.
Further, we notice that $P(\phi)$ of the individual sublattices have
	 smooth single-peaked shapes. Furthermore, all distributions are discrete, reflecting that the number of possible transitions (and thus, $\phi$ values) is finite, because of the discreteness of the underlying spin dynamics. For the colder sublattice, $\mathcal{L}_2$, 
	 $\phi$ only takes a particularly small number of values. This is due to the fact that at low bath temperatures, the sublattice only explores a small part of the phase space, and hence, the number of distinct state transitions is small. 
	For the hotter sublattice, $\mathcal{L}_1$, we notice that the maxima and mean values of $P(\phi)$ lie at $\phi < 0$ [in both cases, (a) and (b)]. This alone would violate the second-law of thermodynamics since it implies a negative mean entropy production rate. However, in its usual form, $\Phi = \Pi>0$, the second law only applies to the entire system which consists of two sublattices, and the negative mean value simply reflects the heat flows from the hotter to the colder heat bath (overall, the entropy is increased over time).

Next, we study the system size dependency of the total entropy production rate (per spin), $\pi$, around the critical point of the phase transition. To this end, we consider a system where $T_2$ is fixed to a value below $T_c^{eq}$ [see the left panel of Fig.~7(a)], and another one where $T_2>T_c^{eq}$ [see the right panel of Fig.~7(a) which is essentially an enlarged version of Fig.~5(a) for different $L$ close to $T_c$]. Both parts of Fig.~7(a) indicate that $\pi$ is \textit{identical} for all system sizes for $T$ values far away from $T_{c}$, i.e., all lines collapse on a single curves. In contrast, the lines split up around $T_{c}$, thus, around the phase transition $\pi$ depends on the system size $L$. This resembles the behavior of the specific heat [recall Fig.~4]. One further observes the emergence of a shoulder which gets more pronounced while increasing $L$. It is, however, noteworthy that we do not observe the formation of a saddlepoint or even non-monotonous behavior for all considered system sizes, i.e., until the value $L=96$.

\begin{figure}
	\includegraphics[width=0.99\linewidth]{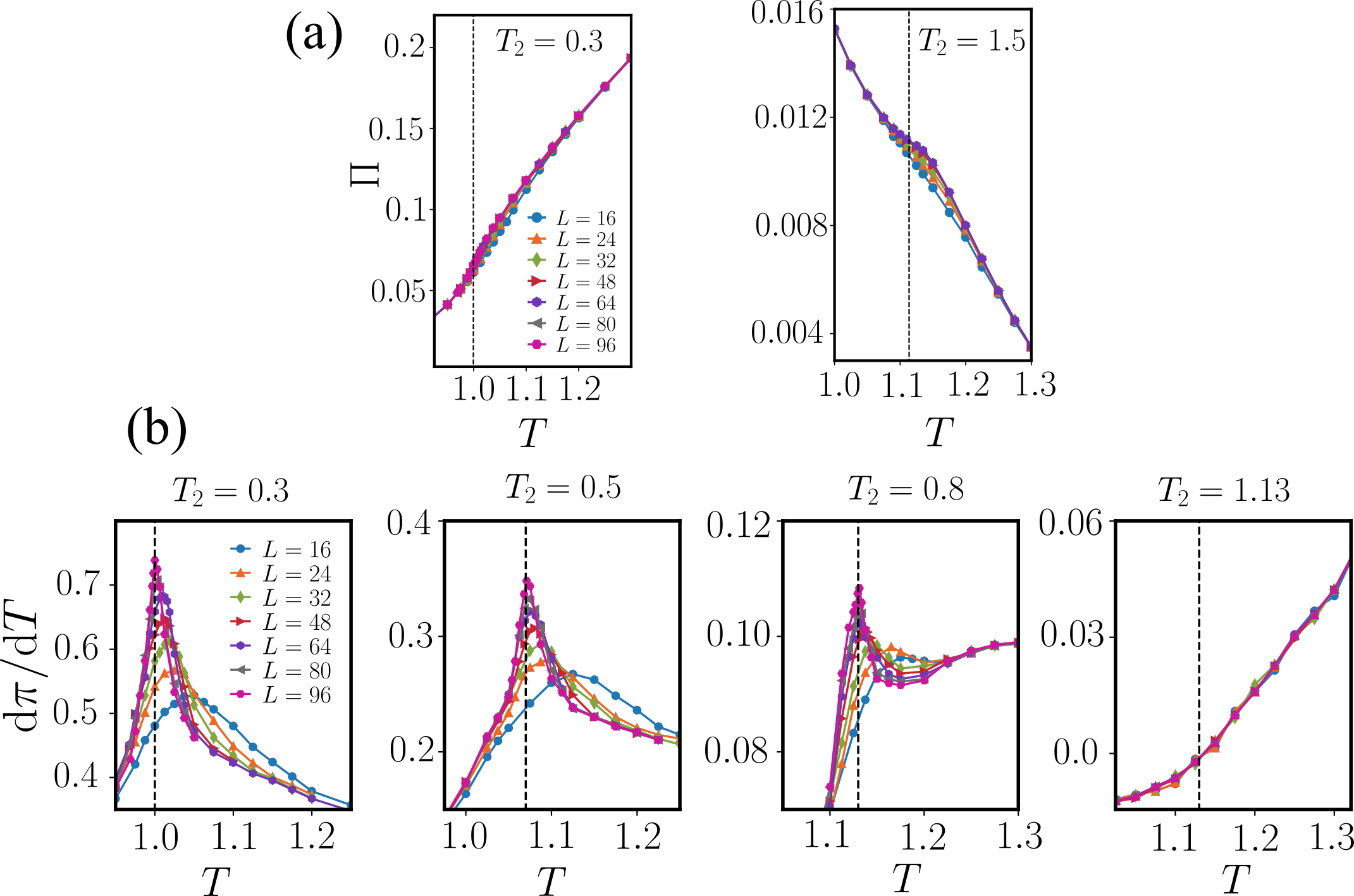}
	\caption{\label{Average cluster shape} $(a)$ The entropy production rate per spin, $\pi$, as function of the mean temperature $T$ for fixed $T_{2}$ and system sizes ranging from $L = 16$ to $L = 96$. In the left panel, the temperature of sublattice $\mathcal{L}_2$ is fixed to $T_{2} = 0.3$ which is below the critical temperature $T_c^{eq}$ of the equilibrium model, while in the right panel, the temperature of $\mathcal{L}_2$ is $T_{2} = 1.5$, which is above $T_c^{eq}$ [see also Fig.~5(a)]. The black dashed lines mark the critical temperature $T_c$. $(b)$ Derivative $\mathrm{d}\pi / \mathrm{d}T$ of the EP rate as function of the mean temperature $T$ for different fixed values of $T_2$ from $T_2 = 0.3$ up to $T_2 = 1.13 = T_c^{eq}$ and system sizes ranging from $L = 16$ to $L = 96$. The black dashed lines mark the critical temperature $T_c$.}
\end{figure}

In order to study the behavior around the critical point, we inspect the derivative of the entropy production rate $\mathrm{d}\pi / \mathrm{d}T$ for various values of $T_{2}$ see Fig.~7. Interestingly, $\mathrm{d}\pi / \mathrm{d}T$ peaks around the temperature of the phase transition. An exception is the case $T_2=T_{c}^{eq}$ where the total EP naturally vanishes and thus does not peak. Moreover, one observes a dependency of the maximum of $\mathrm{d}\pi / \mathrm{d}T$ on the value of $T_2$ which (for fixed $L$) decreases as $T_2$ approaches $T_c^{eq}$. 

\begin{figure}
	\includegraphics[width=0.99\linewidth]{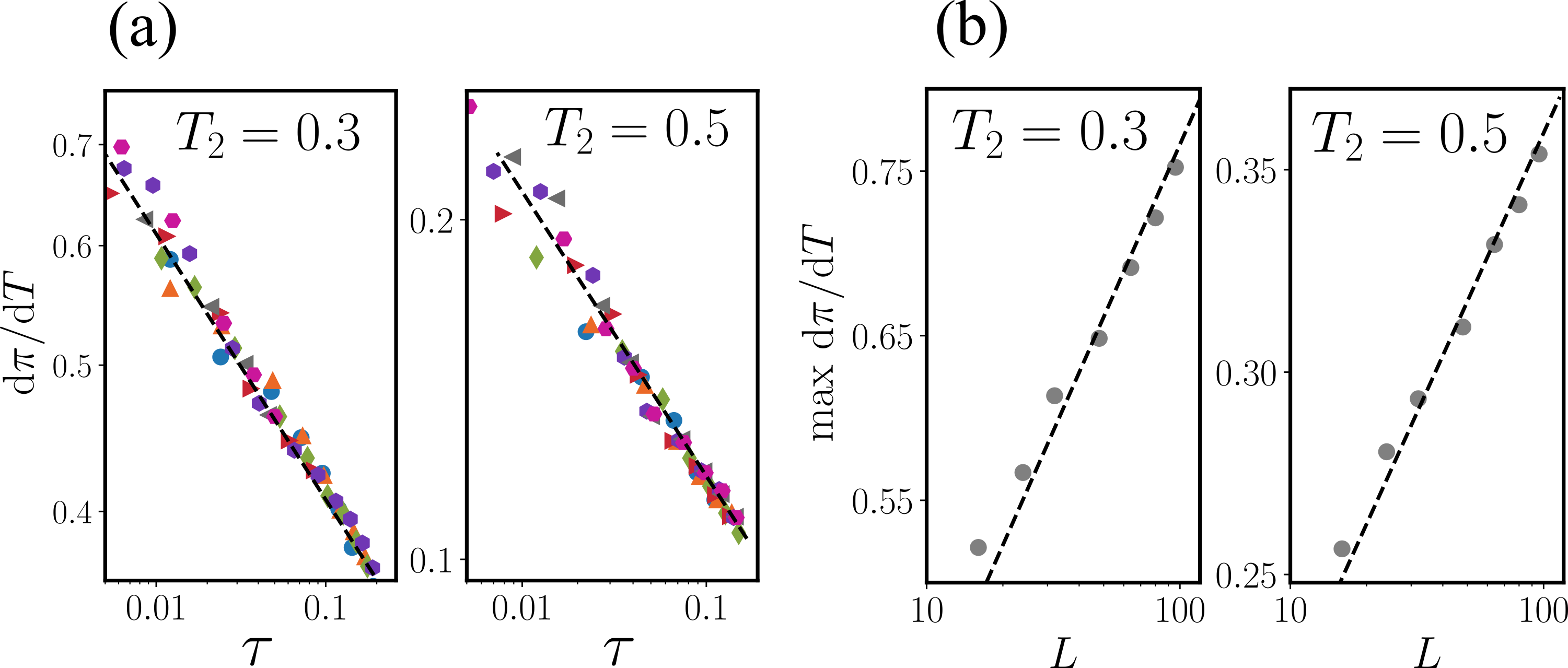}
	\caption{(a) Power-law scaling of the derivative of the entropy production rate as function of the reduced temperature $\tau = |1 - T/T_c|$ for two values of $T_2$ ($T_{2} = 0.3$ and $T_{2} = 0.2$) and system sizes ranging from $L = 16$ to $L = 96$.  The black dashed line in the left panel follows $\sim -0.175(11)$, while in the right panel it follows $\sim -0.145(15)$. (b) Maximum of the derivative of the entropy production rate as function of system size $L$. The left panel shows the scaling of $\mathrm{d}\pi / \mathrm{d}T_{max}$ at $T_2 = 0.3$ for system sizes from $L = 16$ up to $L = 96$. The black dashed lines scales $\sim 0.245$. In the right panel the same is plotted for $T_2 = 0.5$ and the black dashed line follows $\sim 0.205$.}
\end{figure}

To analyze the nonequilibrium phase transitions in more detail, we perform a finite-size scaling, similar to our investigation of the specific heat (see Fig.~4). We aim to stress that the application of a finite-size scaling analysis to the EP at a nonequilibrium transition is, to our knowledge, novel. First, we study the scaling behavior of $\mathrm{d}\pi / \mathrm{d}T$ in the disordered phase as function of the reduced temperature $\tau$. Second, we consider the peak height as function of the system size $L$. As can be seen in Fig.~8(a), $\mathrm{d}\pi / \mathrm{d}T$ shows power-law behavior $\sim \tau^{\zeta}$ with an exponent $\zeta$, whose precise value depends on the distance from equilibrium at the phase transition (i.e., on the value of $\Delta T = |T_2-T_1|$). Specifically, we detect power-law behavior of $\mathrm{d}\pi / \mathrm{d}T$ for all considered values of $T_2$ with a decreasing value for $\zeta$ as $T_2$ approaches $T_c^{eq}$, where it nullifies. For $T_2 = 0.3$ [see the left panel in Fig.~8(a)] the exponent reads $\zeta = 0.175(11)$, while for $T_2 = 0.5$ [see  the right panel in Fig.~8(a)] $\zeta = 0.145(15)$ (see the dashed black lines). While the power-law behavior resembles that of the specific heat, there is a marked difference in the sense that the exponent $\zeta$ is not constant (such as the exponent $\alpha$ of $C_v$), but depends on $\Delta T$. In addition, we analyze the scaling behavior of the maximum of $\mathrm{d}\pi / \mathrm{d}T$ as the system size $L$ is increased and show results for $T_2 = 0.3$ and $T_2 = 0.5$ in Fig.~8(b). According to the finite-size scaling theory for equilibrium systems \cite{Binder_1981}, all divergent quantities scale as $\sim L^{\mathrm{a} / \nu}$, where $\mathrm{a}$ is the critical exponent of the power-law decay of that very quantity. Thus, we test whether the maximum of $\mathrm{d}\pi / \mathrm{d}T$ scales as $\sim L^{\zeta / \nu}$, with $\nu = 2/3$. From our numerical data, we find $\mathrm{d}\pi / \mathrm{d}T_{max} \sim L^{0.245}$ for $T_2 = 0.3$ and $\mathrm{d}\pi / \mathrm{d}T_{max} \sim L^{0.205}$ for $T_2 = 0.5$ which is indeed in good agreement with $\zeta = 0.175(11)$ ($T_2 = 0.3$) and $\zeta = 0.145(15)$ ($T_2 = 0.5$) as obtained in Fig.~8(a). The fulfillment of the finite-size scaling relation shows indeed that the derivative of the entropy production rate behaves as a diverging quantity as the critical point of the phase transition is approached. It further demonstrates that the finite-size scaling theory is applicable to the EP rate, despite the dependency of the critical exponent on the temperature gradient between the two sublattices.

\subsection{BKT-like phase transition in the continuous vector Potts model with $q \rightarrow \infty$}

\begin{figure}
	\includegraphics[width=0.95\linewidth]{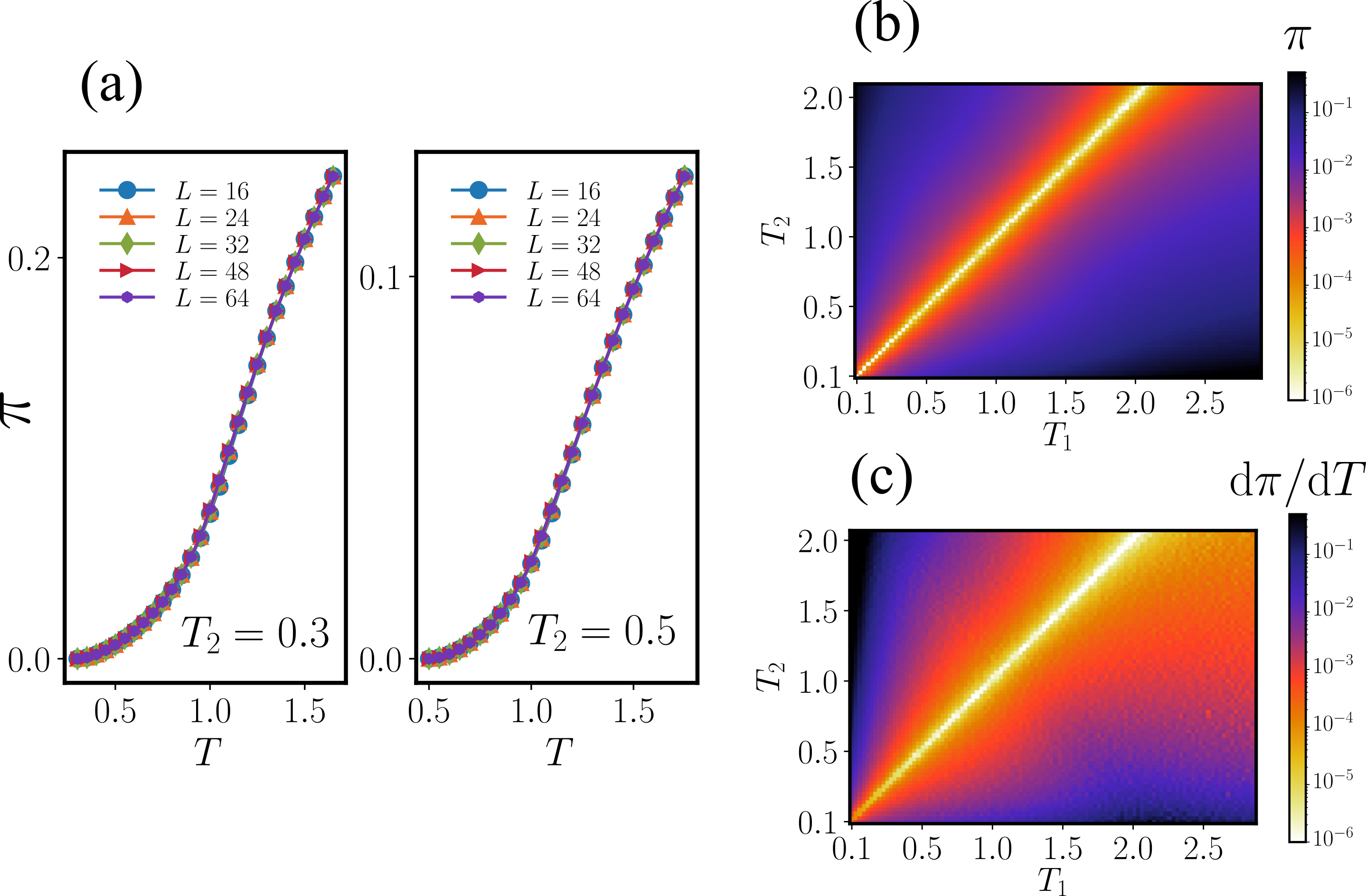}
	\caption{(a) EP rate per spin $\pi$ of the nonequilibrium vector Potts model with $q \rightarrow \infty$ (XY model) as function of the mean temperature $T$ for system sizes ranging from $L = 16$ to $L = 64$ with $T_2 = 0.3$ and $T_2 = 0.5$. (b) Heatmap of $\pi$ in the XY model on a lattice of size $L = 32$ for temperatures of the two sublattices ranging from $T_1 = T_2 = 0.1$ up to $T_1 = 2.0$ and $T_2 = 2.5$. (c) Derivative of the EP rate per spin, $\mathrm{d}\pi / \mathrm{d}T$ in the XY model on a lattice of size $L = 32$.}
\end{figure}

Now we turn to the vector Potts model with $q \rightarrow \infty$ (also known as the XY model), where the spins can freely rotate in the $x-y$ plane, i.e., all spin orientations $\sigma_i \in [0,2 \pi]$ are allowed. As a consequence of the continuous spin symmetry and the two-dimensional character of the system, there exists no long-range ordered phase at finite temperatures as stated by the Mermin-Wagner theorem \cite{Wagner_1966}. Instead, a quasi-long range ordered phase, the BKT phase, occurs at low bath temperatures. While the infinite-order transition between the disordered and the BKT phase is quite well understood in the equilbrium model \cite{Kosterlitz_1974}, nonequilibrium BKT phase transitions are in general less understood. In particular, the question of how the EP rate behaves at this transition has, to the best of our knowledge, not been considered in earlier literature. In the previous discussion of the case $q=4$, we have seen that the derivative of the total EP shows critical behavior which partially resembles the behavior of the specific heat. Let us now see if this analogy carries over to the BKT transition, which has very different overall characteristics and, in particular, is not accompanied with a divergence of $C_\nu$ at the critical temperature which is given by $T_c^{eq} = 0.892880(6)$~\cite{Komura_2012} in the equilibrium XY model~\cite{Kosterlitz_1974,Himbergen_1981,Hasenbusch_2005,Hasenbusch_2008,Komura_2012,Hsieh_2013}. In Fig.~9(a), we show results for $\pi$ at $T_2 = 0.3$ and $T_2 = 0.5$ for system sizes ranging from $L = 16$ up to $L = 64$. As indicated there, the EP rate does not split with respect to $L$ in the vicinity of the phase transition. Instead, $\pi$ is apparently size-independent in the depicted temperature range which includes the BKT transition. In order to visualize the EP rate for different combinations of $T_1$ and $T_2$, we plot $\pi$ in the $T_1-T_2$ plane in Fig.~9(b) together with the derivative of the EP rate with respect to temperature, $\mathrm{d}\pi / \mathrm{d}T$ in Fig.~9(c) for system size $L = 32$.

\begin{figure}
	\includegraphics[width=0.95\linewidth]{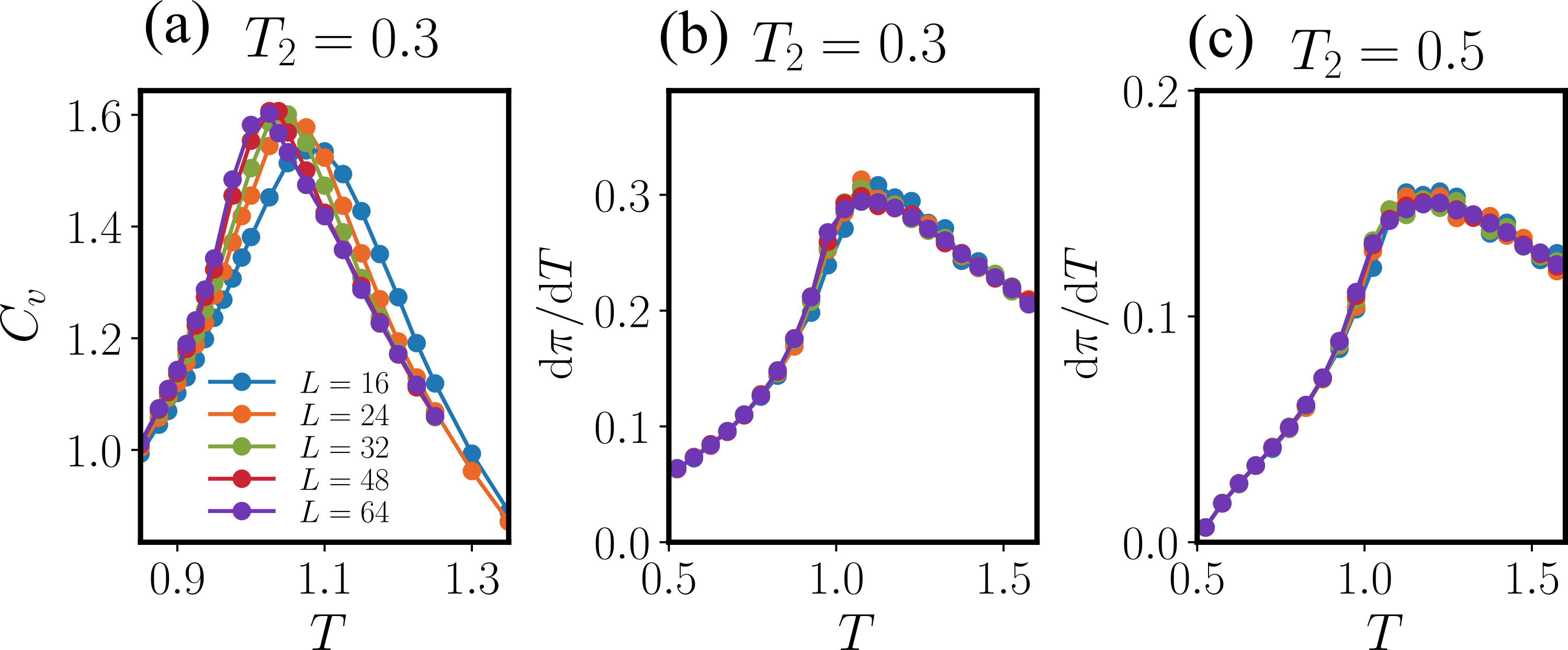}
	\caption{ $(a)$ Specific heat $C_v$ of the nonequilibrium vector Potts model with $q \rightarrow \infty$ (XY model) as function of the mean temperature $T$ for $T_2 = 0.5$ and system sizes ranging from $L = 16$ to $L = 64$. $(b)$ shows the derivative, $\mathrm{d}\pi / \mathrm{d}T$, of the EP rate as function $T$ for the same system sizes and $T_2 = 0.3$, while $T_2 = 0.5$ in $(c)$.}
\end{figure}

Additionally, $C_v$ for $T_2 = 0.5$ and $\mathrm{d}\pi / \mathrm{d}T$ for $T_2 = 0.3$ and $T_2 = 0.5$ are plotted in Fig.~10. In contrast to the PM to FM transition of the $4$-state vector Potts model, $C_v$ in the nonequilibrium XY model does not show any feature like a divergence at criticality. In particular, it only shows a peak around $T=1.1$, as does the equilibrium XY model \cite{Himbergen_1981}, which is above $T_c$. Interestingly, also the derivative of the EP rate with respect to temperature, $\mathrm{d}\pi / \mathrm{d}T$, does not peak in the vicinity of the critical point. Similar to the specific heat, $\mathrm{d}\pi / \mathrm{d}T$ also shows the peak around $T = 1.1$ which does not depend on $L$, i.e., the maximum of $\mathrm{d}\pi / \mathrm{d}T$ does not diverge, but remains constant for all considered system sizes. However, we observe that the maximum of $\mathrm{d}\pi / \mathrm{d}T$ depends on the temperature difference $|T_2 - T_1|$ between the two sublattices in the vicinity of the peak as confirmed by comparing Fig.~10(b) with Fig.~10(c), where one observes that the maximum value of $\mathrm{d}\pi / \mathrm{d}T$ at $T_2 = 0.3$ is larger compared to $T_2 = 0.5$.

\begin{figure}
	\includegraphics[width=0.95\linewidth]{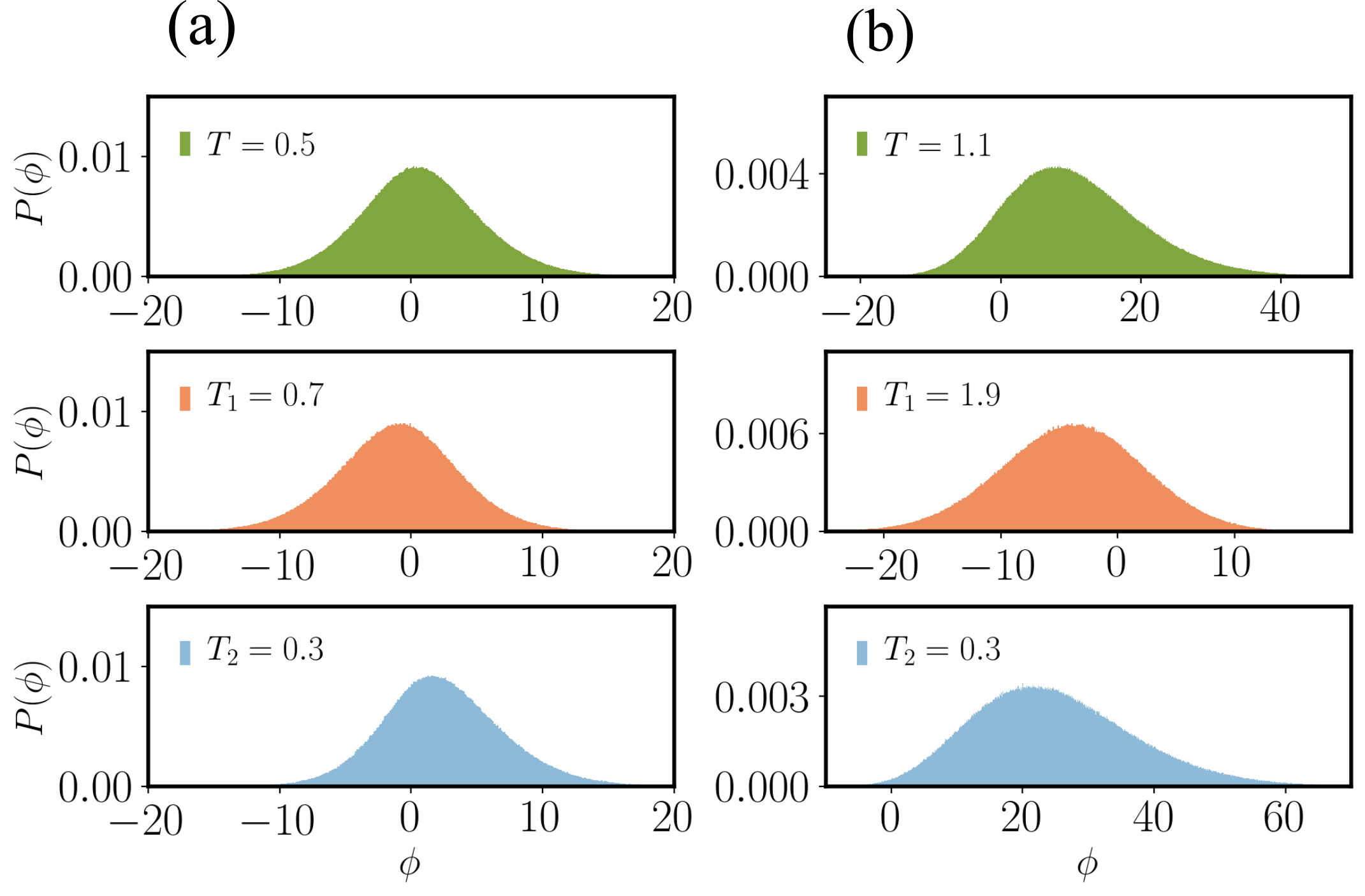}
	\caption{\label{Average cluster shape} Distribution $P(\phi)$ of the medium entropy $\phi = \Delta \phi(l)$ that is produced in the system along stochastic trajectories of length $l = 100$ in the XY model (where $q \rightarrow \infty$). The top panel in $(a)$ shows $P(\phi)$ below the critical point in the BKT phase for a system with $L = 64$ at $T_1 = 0.7$ and $T_2 = 0.3$. The middle panel in $(a)$ shows $P(\phi)$ for $\mathcal{L}_1$ and the one at the bottom of $(a)$ for $\mathcal{L}_2$. $(b)$ shows the same in the PM disordered phase with for $T_1 = 1.9$ and $T_2 = 0.3$.}
\end{figure}

Just as for the vector Potts model with $q = 4$, we investigate the distribution $P(\phi)$ of entropy $\phi = \Delta \phi(l)$ that is produced along stochastic trajectories of length $l = 100$. To this end, we plot $P(\phi)$ for a system of size $L = 64$ in the quasi long-range ordered BKT phase at $T = 0.5$ with $T_1 = 0.7$ and $T_2 = 0.3$ (i.e., $\Delta T = 0.4$) in the top panel of Fig.~11(a). The distribution for the whole system seems to be symmetric around the peak position of $P(\phi)$ which is located in the positive range, $\phi > 0$ in accordance with the second law of thermodynamics. In contrast, $P(\phi)$ for subsystem $\mathcal{L}_1$ peaks in the negative range, and $P(\phi)$ for subsystem $\mathcal{L}_2$ peaks at a positive value of $\phi$. This difference in the peak positions just reflects the expected entropy flow from the hot to the cold reservoir. Additionally, one observes different skew directions for $P(\phi)$ in the two subsystems. $P(\phi)$ for subsystem $\mathcal{L}_1$ is slightly right-skewed, while $P(\phi)$ in $\mathcal{L}_2$ is a left-skewed distribution. This effect becomes more pronounced for the system in the PM phase [see Fig.~11(b)] where one clearly observes that $P(\phi)$ is skewed in both sublattices. Since the distribution for $\mathcal{L}_2$ is stronger skewed, the distribution for the whole system is also left-skewed.

\section{Conclusions and Outlook}

In this paper, we have analyzed the behavior of various critical quantities and that of the total EP rate around the critical point in a nonequilibrium $q$-state vector Potts model (with $q = 4$ and $q \rightarrow \infty$). The nonequilibrium character results from coupling the spins to two heat baths at different temperatures. Based on this nonequilibrium model, we address several questions: 
Does the type of phase transition and the critical exponents change by driving the system away from equilibrium? 
Does the EP exhibit universal behavior around a continuous phase transition? What happens to the EP in the vicinity of an a infinite-order phase transition?

First, we have investigated the model with $q = 4$ in the vicinity of the second--order phase transition. We found that the critical temperature of the transition decreases as the temperature difference between the two heat baths increases. Moreover, the behavior of the specific heat resembles that of the equilibrium model, i.e., it shows power-law divergence with critical exponents that are independent of the temperature difference. Interestingly, the derivative of the EP rate with respect to temperature behaves, to some extent, similar. It also shows power-law divergence. However, the value of the scaling exponents does depend on the temperature difference and is thus non-universal. Concerning the model with $q \rightarrow \infty$, the specific heat as well as the derivative of the EP rate do not show any noticeable behavior around the infinite--order transition from the PM to the quasi long-range ordered BKT phase. Instead, both quantites have a finite peak at a temperature above the critical temperature, i.e., in the PM phase. As the temperature difference between the heat baths increases, the maximum value of the derivative of the EP rate becomes more pronounced. In total, our results provide evidence that the derivative of the EP behaves like a critical quantity, but, as we report here, is non-universal.

Finally, we aim at pointing out perspectives for future work, starting with some questions directly following from the present work. For the sake of generality one should study and compare the behavior of the specific heat with the EP in other dimensions and for different lattice topologies. Further, although the BKT phase transition is not accompanied by a divergence of thermodynamic quantities, in equilibrium it still obeys characteristic scaling dimensions~\cite{Odor_2004}. A more detailed analysis of this transition in the nonequilbrium model, and, specifically, with respect to the derivative of the EP rate, represents an interesting objective of future research. From a theoretical point of view, it would moreover be worth to think about the connection between EP and specific heat, which seem to behave analogously around criticality, on a fundamental level. 

Furthermore, an interesting novel perspective on the nonequilibrium model considered here is the reinterpretation as a model with non-reciprocal coupling between interacing isothermal spins. To be more specific, a vector-Potts model where interacting spins are coupled among each other with two distinct coupling constants ($J_1= J/T_1$ and $J_2= J/T_2$) and uniform temperature follows the exact same equations of motions as our model (with two temperatures and identical coupling constants $J$). This provides a connection to spin models on directed graphs \cite{Sanchez_2002,Castellano_2009,Dorogovtsev_2008,Dorogovtsev_2002,Schwartz_2002,Lima_2006,Lipowski_2015}, and to the topic of non-reciprocal interactions, which is currently a focus in nonequilibrium statistical mechanics \cite{Ivlev_2015,Loos_2019,Kryuchkov_2018}. It would be interesting to compare the thermodynamic properties of spin systems subjected to different driving mechanisms, e.g., non-reciprocal couplings, temperature gradients, external fields and colored noise. 

\section{Acknowledgements}

This work was funded by the Deutsche Forschungsgemeinschaft (DFG, German Research Foundation) - Projektnummer 163436311 - SFB 910.

\bibliographystyle{iopart-num}
\bibliography{BIB}

\end{document}